\documentclass[3p,times,procedia]{elsarticle}
\usepackage{nupha_ecrc}
\usepackage{calc}
\usepackage{graphicx}
\usepackage{color}
\usepackage{nth}
\usepackage{amsmath}
\usepackage{lineno}
\usepackage{amssymb}
\usepackage[figuresright]{rotating}
\usepackage{subcaption}

\volume{00}

\firstpage{1}

\journalname{Nuclear Physics A}

\runauth{P. Steinbrecher}

\jid{nupha}

\jnltitlelogo{Nuclear Physics A}

\newcommand{\ord}{\mathcal{O}}

\let\oldalign\align
\let\oldendalign\endalign

\renewenvironment{align}
  {\linenomathNonumbers\oldalign}
  {\oldendalign\endlinenomath}

\begin{document}

\begin{frontmatter}
\dochead{XXVIIth International Conference on Ultrarelativistic Nucleus-Nucleus Collisions\\ (Quark Matter 2018)}

\title{The QCD crossover at zero and non-zero baryon densities from Lattice QCD}

\author{Patrick Steinbrecher (for the HotQCD collaboration)}

\address{Fakult\"at f\"ur Physik, Universit\"at Bielefeld, D-33615 Bielefeld, Germany\\ Physics Department, Brookhaven National Laboratory, Upton, NY 11973}

\begin{abstract}
We map out the QCD crossover line $\frac{T_c(\mu_B)}{T_c(0)} = 1 - \kappa_2 \left(
\frac{\mu_B}{T_c(0)} \right)^2 - \kappa_4 \left( \frac{\mu_B}{T_c(0)} \right)^4 + \ord(\mu_B^6)$
for the first time up to $\ord(\mu_B^4)$ for a strangeness neutral system by performing a Taylor
expansion of chiral observables in temperature $T$ and chemical potentials $\mu$. At vanishing
chemical potential, we report a crossover temperature $T_c(0) = (156.5 \pm 1.5)\;\mathrm{MeV}$
defined by the average of several second-order chiral susceptibilities. For a system with thermal
conditions appropriate for a heavy-ion collision, we determined a curvature from the subtracted
condensate as $\kappa_2 = 0.0120(20)$ and from the disconnected susceptibility as $\kappa_2 =
0.0123(30)$. The next order $\kappa_4$ is significantly smaller. We also report the crossover
temperature as a function of the chemical potentials for: baryon-number, electric charge,
strangeness and isospin. Additionally, we find that $T_c(\mu_B)$ is in agreement with lines of
constant energy density and constant entropy density. Along this crossover line, we study
net baryon-number fluctuations and show that their increase is substantially smaller compared to
that obtained in HRG model calculations. Similarly, we analyze chiral susceptibility
fluctuations along the crossover line and show that these are constant. We conclude that no signs
for a narrowing of the crossover region can be found for baryon chemical potential $\mu_B <
250\;\mathrm{MeV}$.
\end{abstract}

\end{frontmatter}


\section{Introduction}

We present results from our study of the crossover of quantum chromodynamics (QCD) in (2+1)-flavor
QCD as a function of the baryon chemical potential $\mu_B$. We base this analysis on
findings~\cite{crossover1,crossover2} that at vanishing chemical potentials strong interaction
matter does not have a genuine phase transition from a gas of hadrons and their resonances (HRG)
to a quark-gluon plasma (QGP). Our goal is to understand up to which baryon chemical potential the
crossover is still analytic. In other words, we are searching for signs of a second order QCD
critical point which would be the start of a genuine first order phase transition line.
\begin{figure}[t]
\begin{center}
\includegraphics[width=0.495\textwidth]{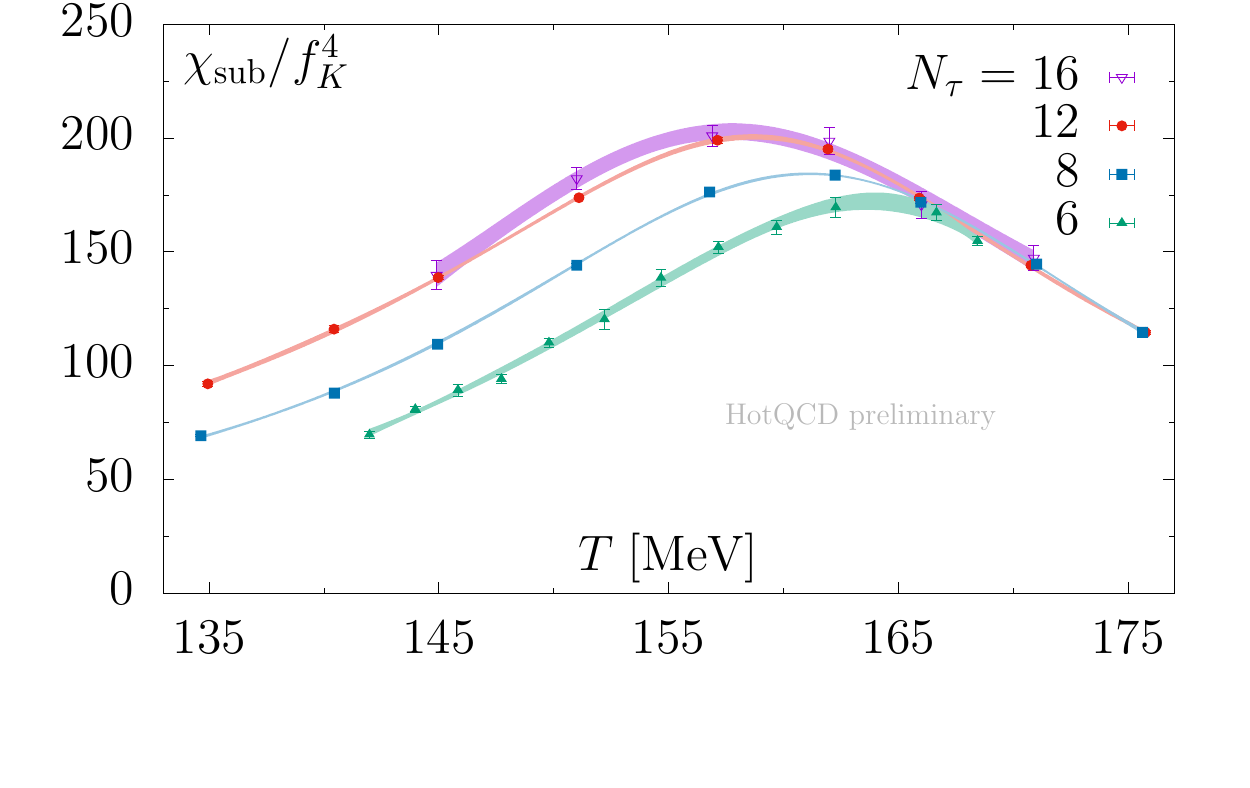}
\includegraphics[width=0.495\textwidth]{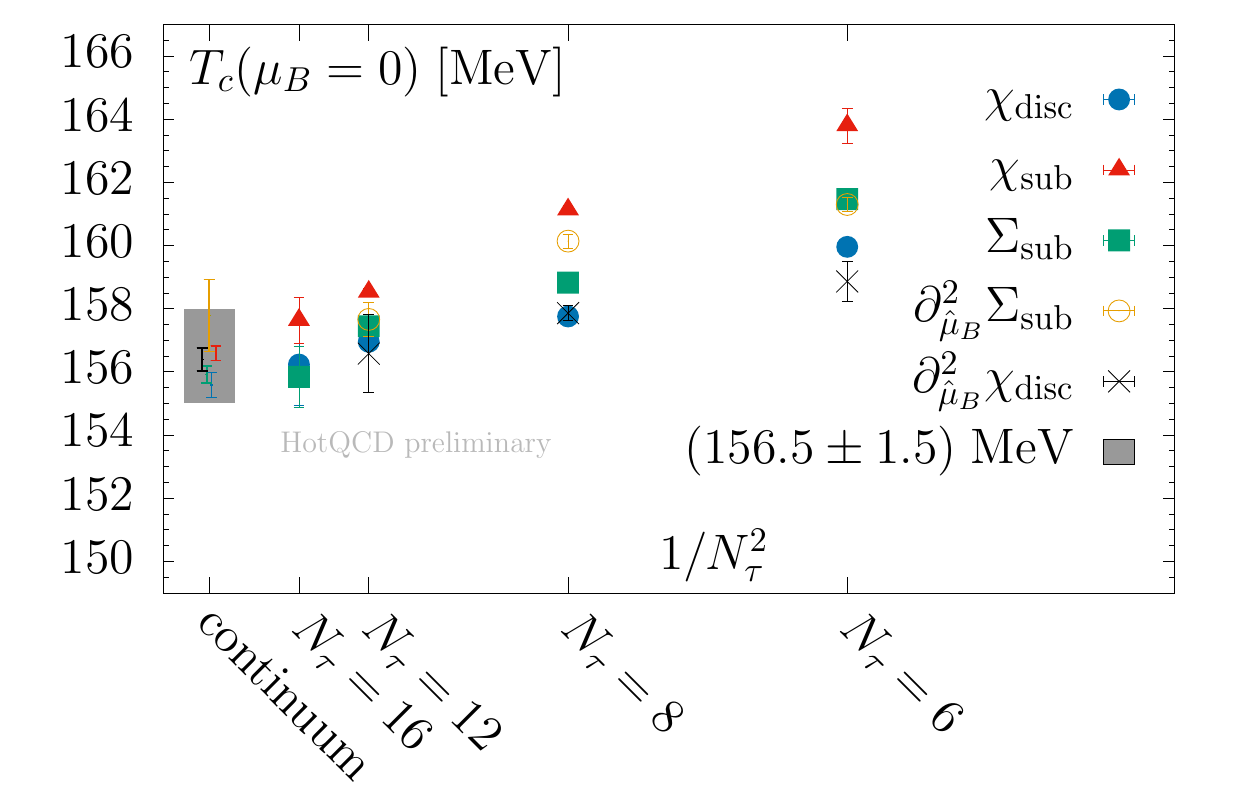}
\vspace{-2.5em}
\end{center}
\caption{The subtracted chiral susceptibility (left) as a function of the temperature for
different $N_\tau$. The data is plotted in two-flavor formulation and normalized using the kaon
decay constant $f_K$. The colored bands are given by AIC weighted Pad\'e approximations which
include statistical and systematic errors. In the right plot, we show the obtained chiral
crossover temperatures $T_c(\mu_B=0)$ as a function of $1/N_\tau^2$ for the subtracted condensate
$\Sigma_\mathrm{sub}$ ($T_c$ defined by inflection point), the subtracted susceptibility
$\chi_\mathrm{sub}$ ($T_c$ defined by maximum), the disconnected susceptibility
$\chi_\mathrm{disc}$ ($T_c$ defined by maximum), the second $\hat\mu_B$ derivative of
$\Sigma_\mathrm{sub}$ ($T_c$ defined by minimum) and the second $\hat\mu_B$ derivative of
$\chi_\mathrm{disc}$ ($T_c$ defined by zero). All chiral observables define pseudo-critical
temperatures. The combined continuum value $(156.5 \pm 1.5)\;\mathrm{MeV}$ in the gray box is an
unweighted average of all observables which includes a $1\;\mathrm{MeV}$ error for setting the
scale. This combined value resembles systematic effects (ambiguity in defining a pseudo-critical
temperature), statistical and scale setting errors.}
\label{fig:chi_obs}
\end{figure}
In the following, we consider chiral observables as the crossover and a possible phase transition
are supposed to be closely related to chiral symmetry restoration. Particularly important are the
subtracted chiral condensate
\begin{equation}
    \Sigma_{\mathrm{sub}} \equiv m_s (\Sigma_u + \Sigma_d) - (m_u+m_d) \Sigma_s
    \qquad\mbox{with}\qquad \Sigma_f = \frac{T}{V} \frac{\partial}{\partial m_f}\ln Z \;,
\end{equation}
the subtracted chiral susceptibility
\begin{equation}
    \chi_\mathrm{sub} \equiv \frac{T}{V} m_s \left(\frac{\partial}{\partial m_u}
                                  + \frac{\partial}{\partial m_d}\right) \Sigma_\mathrm{sub}
\end{equation}
as shown in Fig.~\ref{fig:chi_obs}, and the related disconnected contribution $\chi_{disc}$ to the
total light quark chiral susceptibility. Their Taylor expansions in chemical potentials have been
described in~\cite{myphd}. If a critical point exists, we should be able to observe scaling with
the critical exponents of a three-dimensional Ising model at finite baryon chemical potential.
When approaching a critical point, a significant increase of chiral susceptibility fluctuations
along the crossover must be observed. We have generated gauge field ensembles using a RHMC for 4
lattice volumes with $N_\tau=6,8,12$ and $16$ in a temperature range from $135\;\mathrm{MeV}$ to
$175\;\mathrm{MeV}$. The simulations have been performed using the tree-level improved HISQ
formulation with two degenerate light quarks and a heavier strange quark set to their physical
values corresponding to a pion mass of about $138\;\mathrm{MeV}$. The scale has been set using the
kaon decay constant~\cite{oldtc}.

\section{The QCD crossover line}
The crossover line can be parameterized as
\begin{equation}
    \frac{T_c(\mu_B)}{T_0} = 1 - \kappa_2 \left( \frac{\mu_B}{T_0} \right)^2 - \kappa_4 \left( \frac{\mu_B}{T_0} \right)^4 + \ord(\mu_B^6) \; ,
\end{equation}
where $T_0$ is the crossover temperature at zero chemical potential given by so-called
pseudo-critical temperatures. Their continuum extrapolations are shown in Fig.~\ref{fig:chi_obs}
(right). In the continuum, all considered pseudo-critical temperatures converge to similar
values. This is why we quote a combined value of $T_0 = (156.5 \pm 1.5)\;\mathrm{MeV}$. This
average is in agreement with previous results~\cite{massimotcnew,kappa_katz} obtained with
different lattice formulations. The curvature coefficients $\kappa_n$ can be obtained by requiring
that e.g.\ each order $\mu_B^n$ in $d^2/dT^2 (\Sigma_{\mathrm{sub}}(T,\hat\mu_B)/f_K^4) \equiv 0$
vanishes~\cite{myphd}.
\begin{figure}[t]
\begin{center}
\vspace{-1em}
\includegraphics[width=0.495\textwidth]{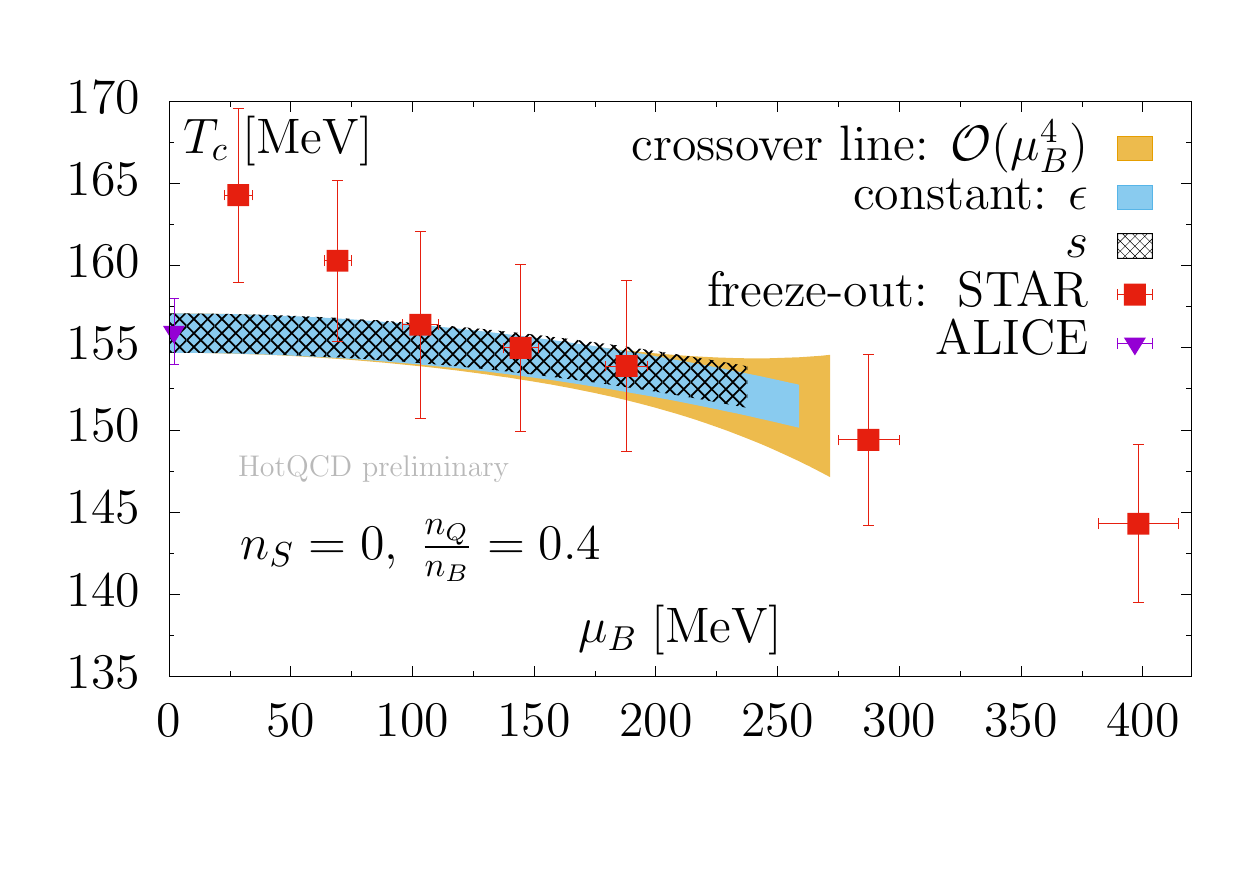}
\includegraphics[width=0.495\textwidth]{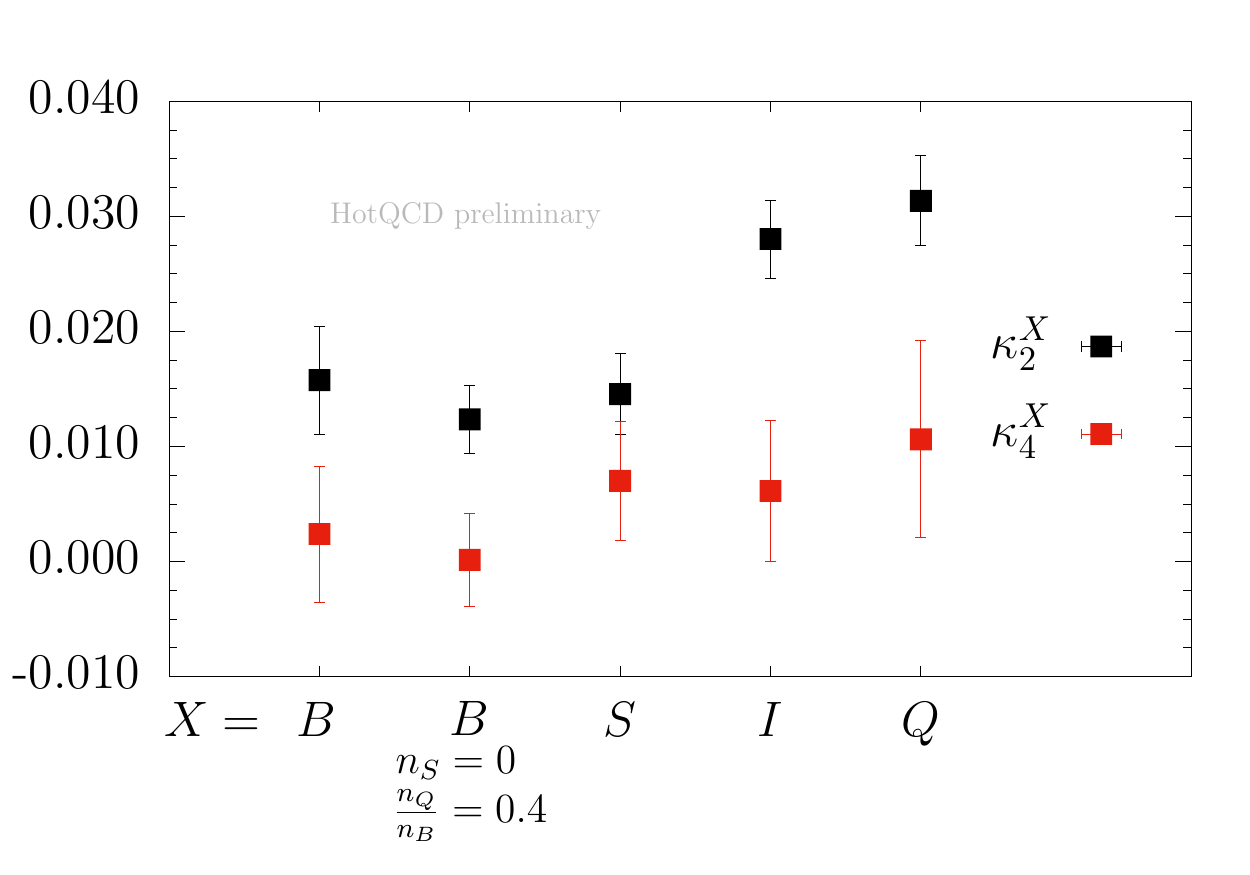}
\vspace{-3.0em}
\end{center}
\caption{The crossover temperature $T_c(\mu_B)$ (left) as a function of the baryon chemical
potential $\mu_B$ for a strangeness neutral system with~$n_Q/n_B = 0.4$ including continuum
extrapolated corrections (yellow band) up to $\ord(\mu_B^4)$. Here, $n_Q$ is the mean electric
charge density and $n_B$ the mean net baryon-number density. All required expansion coefficients
$\kappa_n$ have been determined from the subtracted condensate $\Sigma_\mathrm{sub}$. On top of
it, we show lines of constant physics for energy density $\epsilon$ and entropy density $s$ taken
from~\cite{eosmu6}. The data points represent chemical freeze-out parameters extracted from the
ALICE~\cite{alice} and STAR~\cite{starbes} experiments. The right figure compares the crossover
curvature coefficients $\kappa_2$ and $\kappa_4$ for systems with different constrains. Here, the
crossover line is defined as $T_c(\mu_X)/T_0 = 1 - \kappa_2^X \left( \mu_X/T_0 \right)^2 -
\kappa_4^X \left( \mu_X/T_0 \right)^4 + \ord(\mu_X^6)$ where $T_0$ is the crossover temperature at
zero chemical potentials and $X$ is a placeholder for baryon-number $B$, electric charge $Q$,
strangeness $S$ and isospin $I$. We also show results of the curvature along $\mu_B$ with the
constrains $n_S=0$ and $n_Q/n_B=0.4$. The coefficients have been determined from a Taylor
expansion of $\chi_\mathrm{disc}$. Extracting these coefficients from $\Sigma_\mathrm{sub}$ gives
similar results. The values are listed in~\cite{myphd}.}
\label{fig:crossoverline}
\end{figure}
In Fig.~\ref{fig:crossoverline}, we compare the parameterization of the crossover line with
results on chemical freeze-out temperatures extracted from heavy-ion collision experiments such as
ALICE~\cite{alice} and STAR~\cite{starbes}. The mean of the ALICE freeze-out temperature of
$156(2)\;\mathrm{MeV}$ agrees with our crossover line at almost vanishing baryon chemical
potential. The STAR data seems to extrapolate to a significantly higher freeze-out temperature
resulting in values well above our crossover line which suggests that both cannot hold
simultaneously. However, previous results~\cite{massimotcnew,kappa_katz} report curvatures which
are in agreement with our crossover line. Additionally, we compare the crossover line to lines of
constant physics (LCPs) from lattice QCD simulations~\cite{eosmu6}. The LCP curvatures from energy
density and entropy density agree with the crossover curvature within errors. Furthermore, we
explored the crossover along several directions and for different constrains in the QCD phase
diagram. We found that the QCD phase diagram has very similar curvatures $\kappa_2$ in all
directions except along directions of non-zero electric charge chemical potential $\mu_Q$ and
isospin chemical potential $\mu_I$. In these cases, the curvature $\kappa_2$ is two times larger
(see Fig.~\ref{fig:crossoverline}).

\section{Fluctuations along the QCD crossover}
\begin{figure}[t]
\begin{center}
\includegraphics[width=0.495\textwidth]{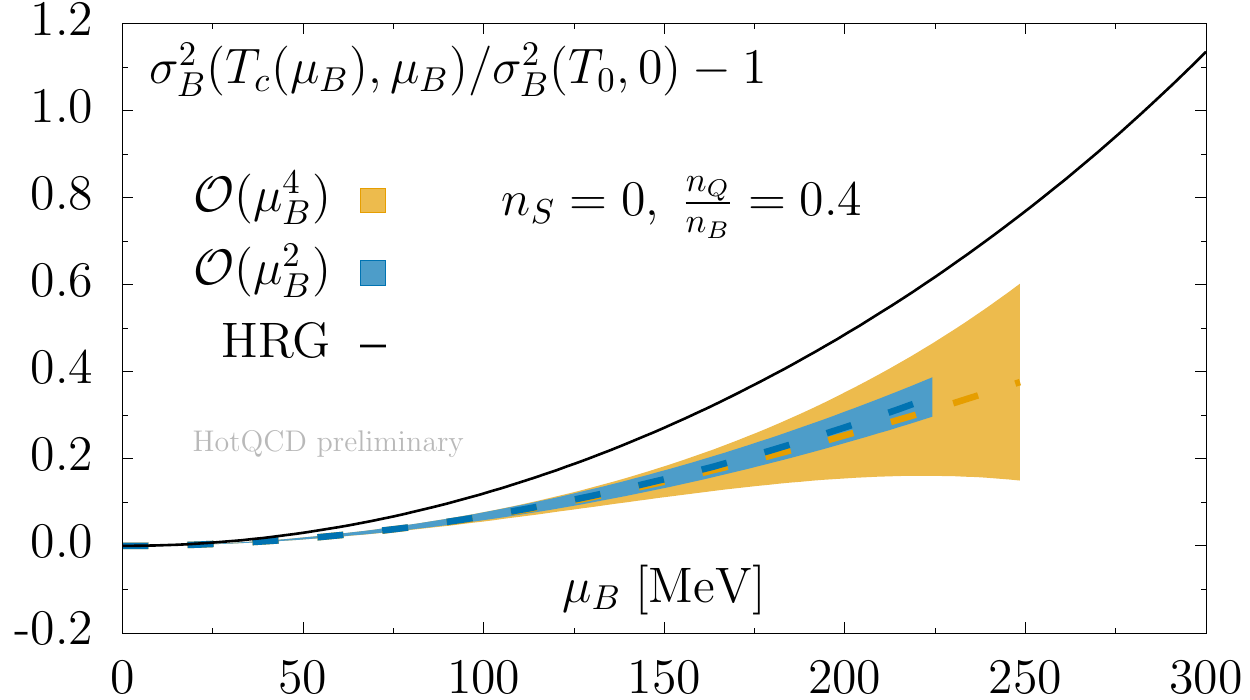}
\includegraphics[width=0.495\textwidth]{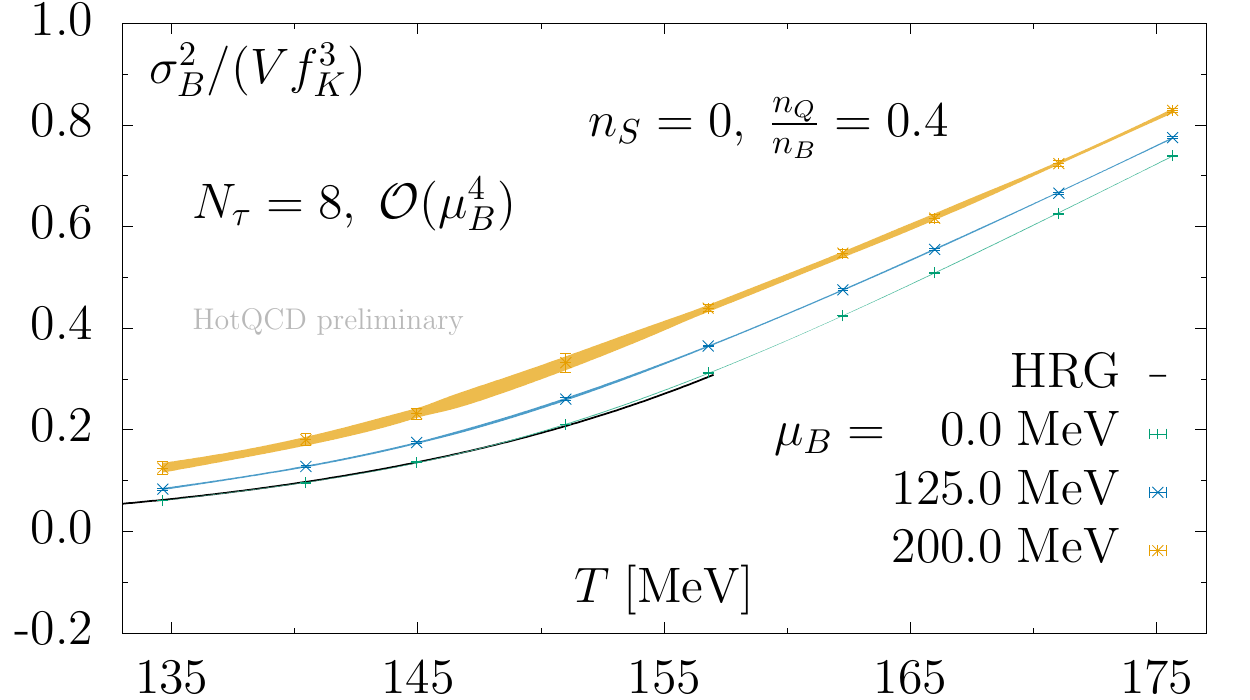}
\vspace{-2.7em}
\end{center}
\caption{The relative change of net baryon-number fluctuations $\sigma^2_B$ (left) along the
crossover line $T_c(\mu_B)$ as a function of $\mu_B$ for a system with strangeness neutrality and
$n_Q/n_B = 0.4$. Here, the curvature of $T_c(\mu_B)$ has been determined from the subtracted
chiral condensate $\Sigma_\mathrm{sub}$. The blue band includes continuum extrapolated corrections
up to $\ord(\mu_B^2)$ and the yellow band up to $\ord(\mu_B^4)$. The corresponding mean is
visualized using a dashed line. Also shown are HRG results using a solid black line evaluated on
along a curvature defined by the mean of $T_c(\mu_B)$. In the right figure, we show $\sigma^2_B$
as function of the temperature at three values of baryon chemical potential $\mu_B$ for a finite
lattice with $N_\tau=8$ including corrections up to $\ord(\mu_B^4)$. For vanishing baryon chemical
potential, we compare QCD results to the HRG as shown by a solid black line.}
\label{fig:baryon_fluc}
\end{figure}
\begin{figure}[t!]
\begin{center}
\vspace{-0.3em}
\includegraphics[width=0.495\textwidth]{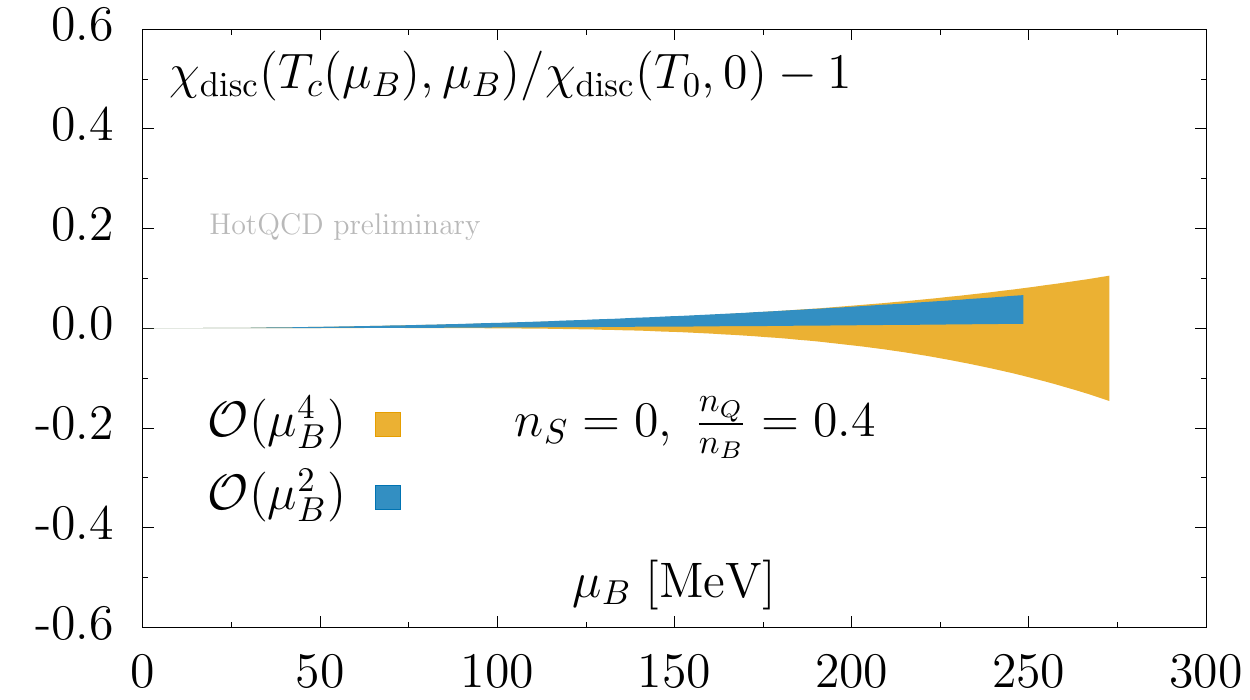}
\includegraphics[width=0.495\textwidth]{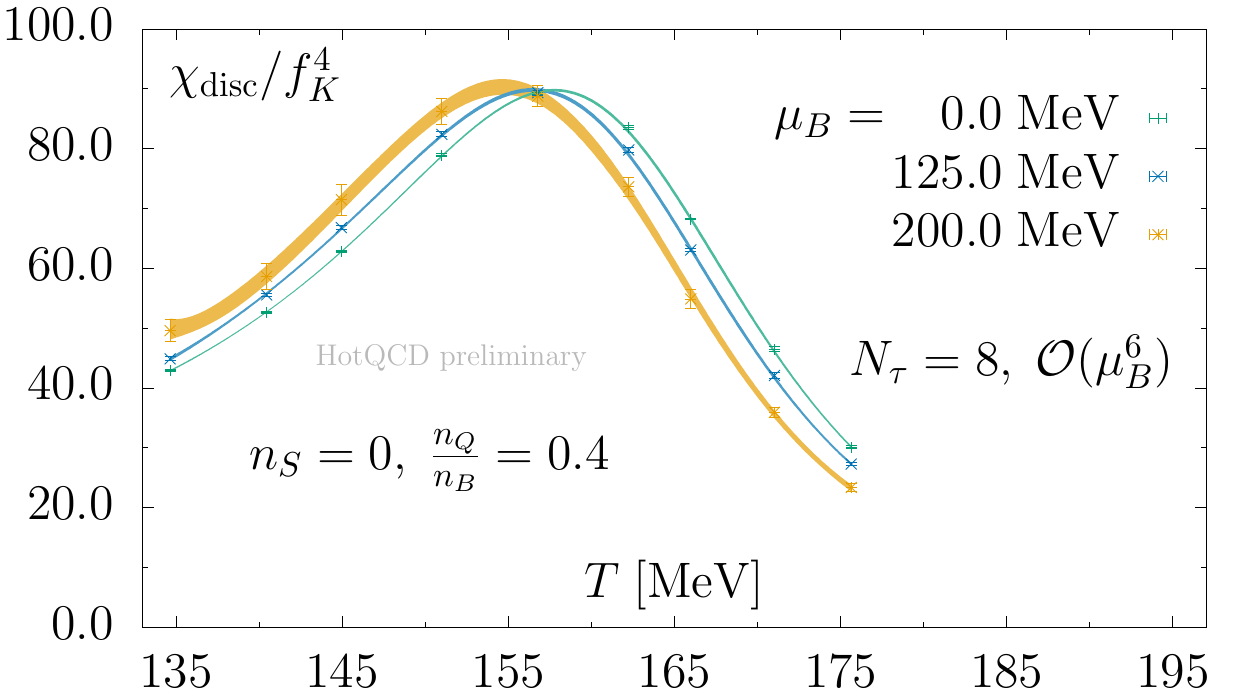}
\vspace{-2.7em}
\end{center}
\caption{The relative change of the disconnected chiral susceptibility
$\chi_\mathrm{disc}$ (left) along the crossover line $T_c(\mu_B)$ as a function of $\mu_B$ for a
system with strangeness neutrality and $n_Q/n_B = 0.4$. Here, the curvature of $T_c(\mu_B)$ has
been determined from $\chi_\mathrm{disc}$. The blue band includes continuum extrapolated
corrections up to $\ord(\mu_B^2)$ and the yellow band up to $\ord(\mu_B^4)$. In the right figure,
we show $\chi_\mathrm{disc}$ as function of the temperature at three values of baryon chemical
potential $\mu_B$ for a finite lattice with $N_\tau=8$ including corrections up to
$\ord(\mu_B^6)$.}
\label{fig:chi_disc_fluc}
\vspace{-0.5em}
\end{figure}
In the following, we study fluctuations of net baryon-number given by
\begin{equation}
    \sigma^2_B = \frac{\partial \ln Z}{\partial \hat\mu_B^2}\;.
\end{equation}
It has been shown successfully~\cite{stephanov1} that net baryon-number fluctuations couple to the
condensate and thus would reveal critical behavior when approaching a critical point. Particularly
interesting is to study their deviations from the HRG model. Even for a finite volume, as given in
heavy-ion collisions, these fluctuations should resemble some critical behavior in the vicinity of
a critical point, i.e.\ show substantially larger fluctuations compared to a HRG. The relative
change of $\sigma_B^2$ can be expressed in a Taylor series
\begin{equation}
    \frac{\sigma_B^2(T_c(\mu_B),\mu_B) - \sigma_B^2(T_0,0)}{\sigma_B^2(T_0,0)} = \lambda_2 \left( \frac{\mu_B}{T_0}\right)^2 + \lambda_4 \left( \frac{\mu_B}{T_0}\right)^4 + \,\ord(\mu_B^6) \;,
\end{equation}
where the expansion coefficients $\lambda_n$ can be determined using lattice QCD. We have
continuum extrapolated these coefficients up to $\ord(\mu_B^4)$ which are used in
Fig.~\ref{fig:baryon_fluc} to visualize the relative change along $T_c(\mu_B)$. For the
strangeness neutral case, the fluctuations are at least a factor two smaller compared to a HRG.
Given that Taylor expansions for baryon-number fluctuations in the HRG model have an infinite
radius of convergence and substantially larger fluctuations compared to our lattice results, we
conclude that it is unlikely that a QCD critical point can be found for $\mu_B <
250\;\mathrm{MeV}$ along the crossover line. Similarly, we have studied chiral susceptibility
fluctuations along the crossover line. As can be seen from Fig.~\ref{fig:chi_disc_fluc}, this
analysis shows a constant peak height for $\chi_\mathrm{disc}$, i.e.\ no significant change along
the crossover line. This suggests that for $\mu_B<250\;\mathrm{MeV}$ no signs for a narrowing of
the crossover region or increasing correlation length have been observed. In addition, we also
measured \nth{6} order expansion coefficients for a fixed lattice spacing and found that these
higher order corrections are negligible for $\mu_B < 250\;\mathrm{MeV}$ for $\chi_\mathrm{disc}$
and $\sigma_B^2$ along the crossover line.

\paragraph{Acknowledgments} This work was supported through Contract No.\ DE-SC001270 with the
U.S.\ Department of Energy and through the grant CRC-TR 211 ``Strong-interaction matter under
extreme conditions'' with the Deutsche Forschungsgemeinschaft (DFG). Numerical calculations have
been made possible through computing resources granted by ALCC, INCITE, NESAP, PRACE and USQCD.

\bibliographystyle{elsarticle-num}
\bibliography{qm18_proc}

\begin{thebibliography}{10}
\expandafter\ifx\csname url\endcsname\relax
  \def\url#1{\texttt{#1}}\fi
\expandafter\ifx\csname urlprefix\endcsname\relax\def\urlprefix{URL }\fi
\expandafter\ifx\csname href\endcsname\relax
  \def\href#1#2{#2} \def\path#1{#1}\fi

\bibitem{crossover1}
Y.~Aoki, et~al., Nature 443 (2006) 675--678.
\newblock \href {http://arxiv.org/abs/0611014} {\path{arXiv:0611014}}.

\bibitem{crossover2}
T.~Bhattacharya, et~al., Phys. Rev. Lett. 113~(8) (2014) 082001.
\newblock \href {http://arxiv.org/abs/1402.5175} {\path{arXiv:1402.5175}}.

\bibitem{myphd}
P.~Steinbrecher, \url{pub.uni-bielefeld.de/publication/2919977} (2018).

\bibitem{oldtc}
A.~Bazavov, et~al., Phys. Rev. D 85 (2012) 054503.
\newblock \href {http://arxiv.org/abs/1111.1710} {\path{arXiv:1111.1710}}.

\bibitem{massimotcnew}
C.~Bonati, et~al., (2018), arXiv:1805.02960.

\bibitem{kappa_katz}
R.~Bellwied, et~al., Phys.Lett. B 751 (2015) 559--564.
\newblock \href {http://arxiv.org/abs/1507.07510} {\path{arXiv:1507.07510}}.

\bibitem{eosmu6}
A.~Bazavov, et~al., Phys.Rev. D 95 (2017) 054504.
\newblock \href {http://arxiv.org/abs/1701.04325} {\path{arXiv:1701.04325}}.

\bibitem{alice}
M.~Floris, Nucl. Phys. A 931 (2014) 103.
\newblock \href {http://arxiv.org/abs/1408.6403} {\path{arXiv:1408.6403}}.

\bibitem{starbes}
{STAR Collaboration}, Phys.Rev. C 96 (2017) 044904.
\newblock \href {http://arxiv.org/abs/1701.07065} {\path{arXiv:1701.07065}}.

\bibitem{stephanov1}
Y.~Hatta, M.~A. Stephanov, Phys. Rev. Lett. 91 (2003) 102003.
\newblock \href {http://arxiv.org/abs/0302002} {\path{arXiv:0302002}}.

\end{thebibliography}

\end{document}